\documentstyle[times,psgreek,namedreferences,kaprotat,epsf]{spackap} 

\begin{opening}

\title{Lithium Depletion in the Sun: 
       A Study of \\ Mixing Based on Hydrodynamical Simulations}

\author{T.\ \surname{Bl\"ocker}}
\institute{Max-Planck-Institut f\"ur Radioastronomie, Bonn, Germany}
\author{H.\ \surname{Holweger}}
\author{B.\ \surname{Freytag}}
\institute{Institut f\"ur Theoretische Physik und Astrophysik, Universit\"at
           Kiel, Germany}
\author{F.\ \surname{Herwig}}
\institute{Astrophysikalisches Institut Potsdam, Germany}
\author{H.-G.\ \surname{Ludwig}}
\institute{Astronomical Observatory, Copenhagen, Denmark}
\author{M.\ \surname{Steffen}}
\institute{Institut f\"ur Theoretische Physik und Astrophysik, Universit\"at
           Kiel, Germany \\
           Astrophysikalisches Institut Potsdam, Germany}

\date{}

\end{opening}

\runningauthor{T.\ Bl\"ocker et al.}
\runningtitle{Lithium depletion in the Sun}

\begin{document}

\begin{abstract}
Based on radiation hydrodynamics modeling of stellar convection zones,
a diffusion scheme has been devised describing the downward penetration
of convective motions
beyond the Schwarzschild boundary (overshoot) into the radiative
interior. This scheme of exponential diffusive overshoot 
has already been successfully applied to AGB stars.    
Here we present an application to the Sun in order to determine
the time scale and depth extent of this additional mixing, i.e. diffusive
overshoot at the base of the convective envelope. We calculated the 
associated destruction of lithium during the evolution towards and on
the main-sequence. We found that the slow-mixing processes induced by the 
diffusive overshoot may lead to a substantial depletion of lithium during
the Sun's main-sequence evolution. 
\end{abstract}

\section{Introduction}
Since lithium is destroyed already at temperatures of $2.5 \cdot 10^{6}$\,K
by nuclear burning in stellar interiors, its surface abundances can be
considerably affected by a sufficiently deep reaching surface convection
zone.
The solar Li problem is the long-standing conflict between 
the observed photospheric Li depletion of the Sun by 2.15\,dex \cite{AG89}
and the predictions of stellar evolution models based on the 
standard mixing-length prescription. The latter show only moderate 
Li depletion during the pre main-sequence (PMS) phase (0.3--0.5\,dex) whereas
the depletion during the main-sequence evolution is negligible. In contrast,
observations of open clusters indicate that effective Li depletion takes 
place on the main-sequence. Consequently,
in order to account for the observations, at least one additional mixing 
mechanism must operate in the radiative regions below the bottom of the 
surface convection zone. 

Suggested solutions include
{\em mass loss} to expose depleted matter from the interior 
   (e.g.\ Schramm et al.,\,1990),
{\em microscopic diffusion} leading to a leakage of Li out of the
          surface convection zone (e.g.\ Michaud,\,1986),
mixing due to {\em internal gravity waves} arising from pressure
          fluctuations in convective flows (e.g.\ Press,\,1981),
rotationally induced mixing by {\em meridional circulation},
and 
rotationally induced mixing due to {\em shear instabilities}
          associated with differential rotation 
(e.g.\ Zahn,\,1992; or Pinsonneault et al.,\,1992; 
Deliyannis \&  Pinsonneault,\,1997  (``Yale'' models)).

Among G and K stars in general, the current situation appears
still controversial. For example, 
\citeauthor{SBKD97} \shortcite{SBKD97}
conclude that the combined evidence of Li and Be in G and K stars
rules out all but the ``Yale'' models. On the other hand,
\citeauthor{MC96} \shortcite{MC96}
criticize the angular momentum loss law adopted in these models, 
and find that rotation inhibits, rather than enforces, depletion.

For more details on the question of Li depletion see, e.g., 
\citeauthor{P97} \shortcite{P97} or \citeauthor{C98} \shortcite{C98} and
references therein.

\section{Our approach to the solar Li problem}
Based on radiation hydrodynamics modeling of stellar convection zones,
a diffusion scheme has been devised describing the mixing process due to 
the downward penetration of convective flows into the radiative interior
\cite{FLS96}.  

\citeauthor{FLS96} \shortcite{FLS96} have investigated the interface between
stellar surface convection zones and the radiative interior
in favourable cases where modeling of 
the entire convection zone and adjacent overshoot region has been possible: 
white dwarfs and A-type main-sequence stars. 
Below the 
classical overshoot layers (characterized by a well defined (anti)correlation
between velocity and temperature fluctuations), the numerical models 
show an extended region where the rms velocity fluctuations decrease 
exponentially with depth. The existence of this low-amplitude velocity field 
is in the end 
simply a consequence of the conservation of mass. Randomly modulated by 
deep-reaching plumes, the resulting flow gives rise to 
{\em diffusive mixing without significant temperature perturbations}. 
The nature of the underlying velocity field is fundamentally different from 
that of propagating gravity waves: while the latter represent oscillating
motions with amplitudes increasing with depth, the former are the extension 
of closed convective flows decaying into the stable layers (for
details see \citeauthor{FLS96} 1996).
Extended overshoot leads to slow mixing of a total mass 
that can exceed that of the convection zone proper by a large factor. 
It is much more efficient than microscopic diffusion, but otherwise similar. 
The corresponding depth-dependent diffusion coefficient can be derived from the
hydrodynamical models and expressed in terms of an efficiency
parameter $f$, the ratio of the scale heights of rms velocity and pressure. 

\begin{figure}
\centering
\epsfxsize=\textwidth
\mbox{\epsffile{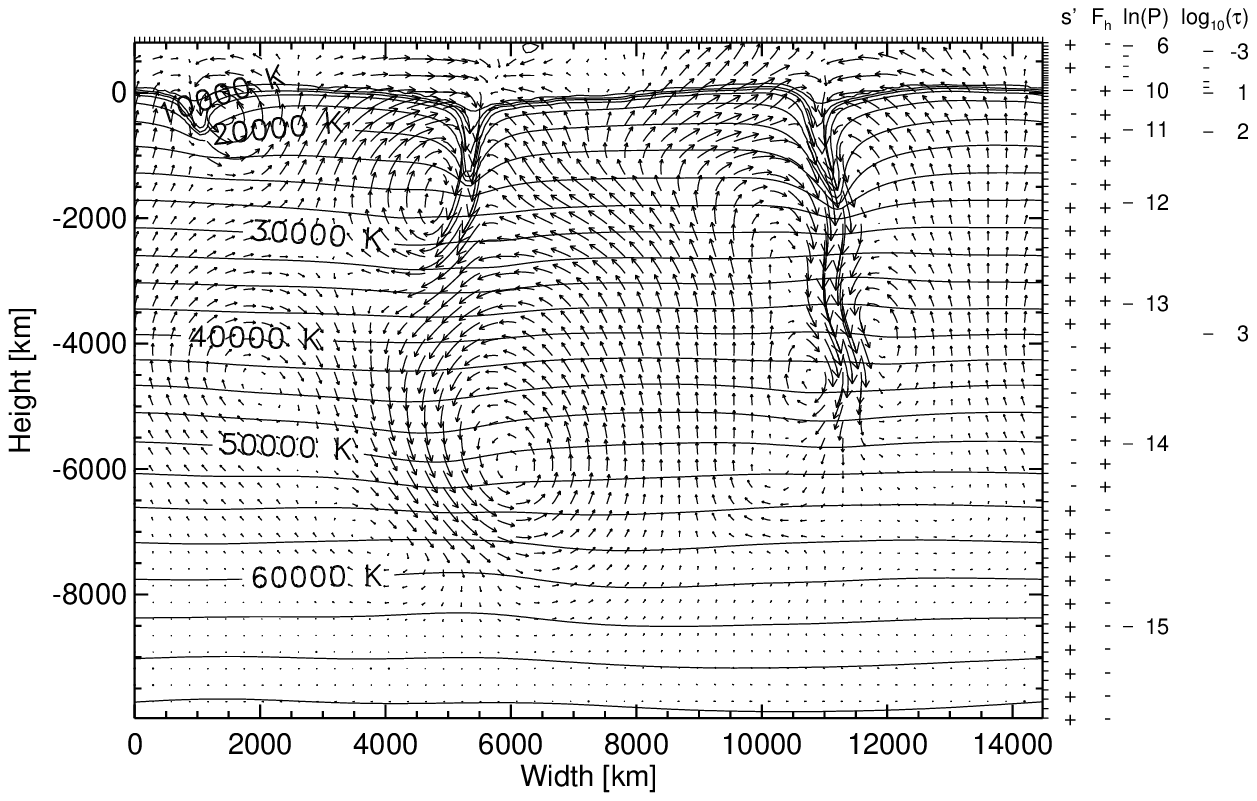}}
\caption{
Snapshot of a
hydrodynamical simulation for an A-type star
with $T_{\rm eff} = 7500$\,K, $\log g = 4.4$ and solar metallicity,
comprising the photosphere and two distinct convection zones:
the upper one is due to combined HI/HeI ionization,
the lower one is produced by HeII ionization.
The velocity field is indicated by pseudo streamlines integrated over
$t_{\rm int} =50$~s (maximum velocity $v_{\rm max}=18.9\,$km\,s$^{-1}$),
contour lines indicate the temperature field in steps of 2500\,K.
Small tick marks at the top and the right show the positions of the
computational grid points ($182 \times 95$).
Four columns at the right show some horizontally averaged quantities:
the sign of the vertical entropy gradient,
the sign of the vertical enthalpy flux,
the natural logarithm of the mean pressure,
and the logarithm of the optical depth.
A negative entropy gradient defines a
convectively unstable region. Note that the stable layer separating both
convection zones (s':+) is completely mixed due to overshoot.
Convection carries up to 30\,\% of the total energy flux.
} \label{Fhydro}
\end{figure}

Fig.~\ref{Fhydro} illustrates the corresponding hydrodynamical simulation of 
the shallow convection zones of an A-type star. Two convective
plumes penetrate deeply into radiative regions: The Schwarzschild 
border of convective instability is located  at an height of 
$\approx -6000$\,km whereas the plumes extend down to -8000\,km, 
corresponding to about one pressure scale height. Convection carries up to
30\,\% of the total energy flux.

In the case of the Sun we are not yet able to include the entire 
convective  envelope in our simulation box and, thus, do not know $f$.
However, we know its approximate value 
for surface convection of  A-type stars, $f \approx 0.25$,
and main-sequence core convection, $f = 0.02$, respectively
(Freytag et al.,\,1996; Herwig et al.,\,1997). 
In the following, we will address the question of
{\em whether this slow mixing, with a ``reasonable'' choice of $f$, can be 
considered a viable alternative to rotational mixing for explaining the 
depletion of Li  during the main-sequence phase of the Sun.}
\section{Application to stellar evolution calculations}
The scheme of exponential diffusive overhoot has already successfully been 
applied in stellar evolution calculations to core and deep envelope convection
\cite{HBSE97}.
Introducing one single efficiency parameter, $f$,
it was possible to account
for the observed width of the main-sequence as well as for important properties
of AGB stars, namely efficient dredge-up processes to produce carbon stars
at low luminosities as required by observations,
and the formation of
$^{13}$C within the intershell region during the thermal pulses 
as the neutron source ($^{13}$C($\alpha$,n)$^{16}$O) for the $s$-process.
Since these calculations deal with the very deep stellar interior,
f was found to be considerably smaller than for the shallow surface convection
of A-type stars, namely 0.02, in accordance with the corresponding 
quasi-adiabatic conditions which allow only small growth rates for 
convective perturbations.
Note, that {\it only} with the inclusion of additional mixing processes 
dredge up was obtained, and that sufficient amounts of  
$^{13}$C are formed {\it only} due to {\it slow} mixing schemes.
For more details, see \citeauthor{HBSE97} \shortcite{HBSE97} or
\citeauthor{B98} \shortcite{B98}.

Abundance changes due to nuclear burning (nuc) and mixing (mix) 
are calculated according to   
\begin{equation}
        \frac{{\rm d}X_i}{{\rm d}t}=\left( \frac{\partial{X_i}}{\partial t} 
        \right)
        _{\rm nuc} + \frac{\partial}{\partial m_{\rm r}}
        \left[ \left(4\pi r^2\rho\right)^2 D 
        \frac{\partial X_i}{\partial m_{\rm r}} \right] _{\rm mix}
\label{diff-gl}
\end{equation}
with $X_{i}$ being the mass fraction of the respective element, 
$m_{r}$ the mass coordinate, and $D$ the diffusion coefficient.
Nuclear burning is treated with a detailed 
nucleosynthesis network.  The choice of $D$ depends on the mixing model.
Within convectively unstable regions according to the Schwarzschild criterion
we follow  \citeauthor{LEF85} \shortcite{LEF85} and 
adopt $D_{\rm conv}= 1/3\,\,v_{\rm c} l$ with 
$l$ being the mixing
length and $v_{\rm c}$ the average velocity of the 
convective elements according to the mixing length theory \cite{BV58}.
The depth-dependent diffusion coefficient of the extended overshoot regions
is given by \citeauthor{FLS96} \shortcite{FLS96}:
\begin{equation} \label{dhyd}
           D_{\rm over} = D_0 \exp{\frac{-2 z}{H_{\rm v}}}, \,\,\,\,\,\,\,
           H_{\rm v} = f \cdot H_{\rm p},  
\end{equation}
where $z$ denotes the distance from the edge of the convective zone 
($z=|r_{\rm edge}-r|$ with $r$: radius), and 
$H_{\rm v}$ is the velocity scale height 
of the overshooting convective elements at $r_{\rm edge}$, 
given as a fraction of the pressure scale height $H_{\rm p}$.
Consequently, $f$ expresses the efficiency of the mixing process.
For $D_{0}$ we take the value of $D_{\rm conv}$ near the 
convective boundary $r_{\rm edge}$. Note that $D_{0}$ is well defined
because $D_{\rm conv}$ drops almost discontinuously at $r_{\rm edge}$.
\section{Model Calculations}\label{modcalc}
{\em Solar models} were evolved from the birthline in the HRD through 
the pre-main-sequence (PMS) and main-sequence phase up to an age of 10 Gyr.
The calculations are based on the code described by 
\citeauthor{B95} \shortcite{B95} and
\citeauthor{HBSE97} \shortcite{HBSE97}.
We use the most recent opacities of 
\citeauthor{IRW92} \shortcite{IRW92} and
\citeauthor{IR96} \shortcite{IR96} complemented with the low-temperature
tables of 
\citeauthor{AF94} \shortcite{AF94}.
With an initial composition of $(Y,Z)=(0.277,0.02)$, we get 
a mixing length parameter of $\alpha = 1.66$ to fit solar
radius and luminosity at $t=4.6$\,Gyr.
At the solar age, the {\em depth\/} of the convection zone is 0.282 $R_\odot$, 
a value slightly lower than the currently adopted helioseismic value of
0.287 $R_\odot$.
%

{\em Convection zones in PMS models} are deep-reaching and massive.
Fig.~\ref{Fmt} illustrates that during the PMS evolution the 
depth of the convection zone changes rapidly in mass 
(compared to the time scales of the main-sequence evolution).
We cannot assume that $f$ is constant during this phase 
and that it has the same value as on the main-sequence.
Thus, an initial ZAMS model was generated by evolving a PMS model
with properly adjusted (mixing) parameters to fit
the `observed' Li depletion of 0.3\,dex in accordance with
results from young open clusters \hbox{(Jones et al.,\,1997).}

The structural 
and nuclear evolution was calculated for a total of ten {\em main-sequence 
$1 M_\odot$ models,} each with a fixed value of $f$
ranging between $f$ = 0.02 and 0.31.
The dependence of structural properties 
of the models on $f$ was found to be negligible.
Apart from taking into account mixing, 
these are standard models and do not include any effects of
rotation, microscopic diffusion, internal gravity waves, accretion, 
magnetic fields,  or mass loss.

\section{Results and discussion}
\begin{figure}[tb]
\centering
\epsfxsize=0.7\textwidth
\rotate[r]{\mbox{\epsffile{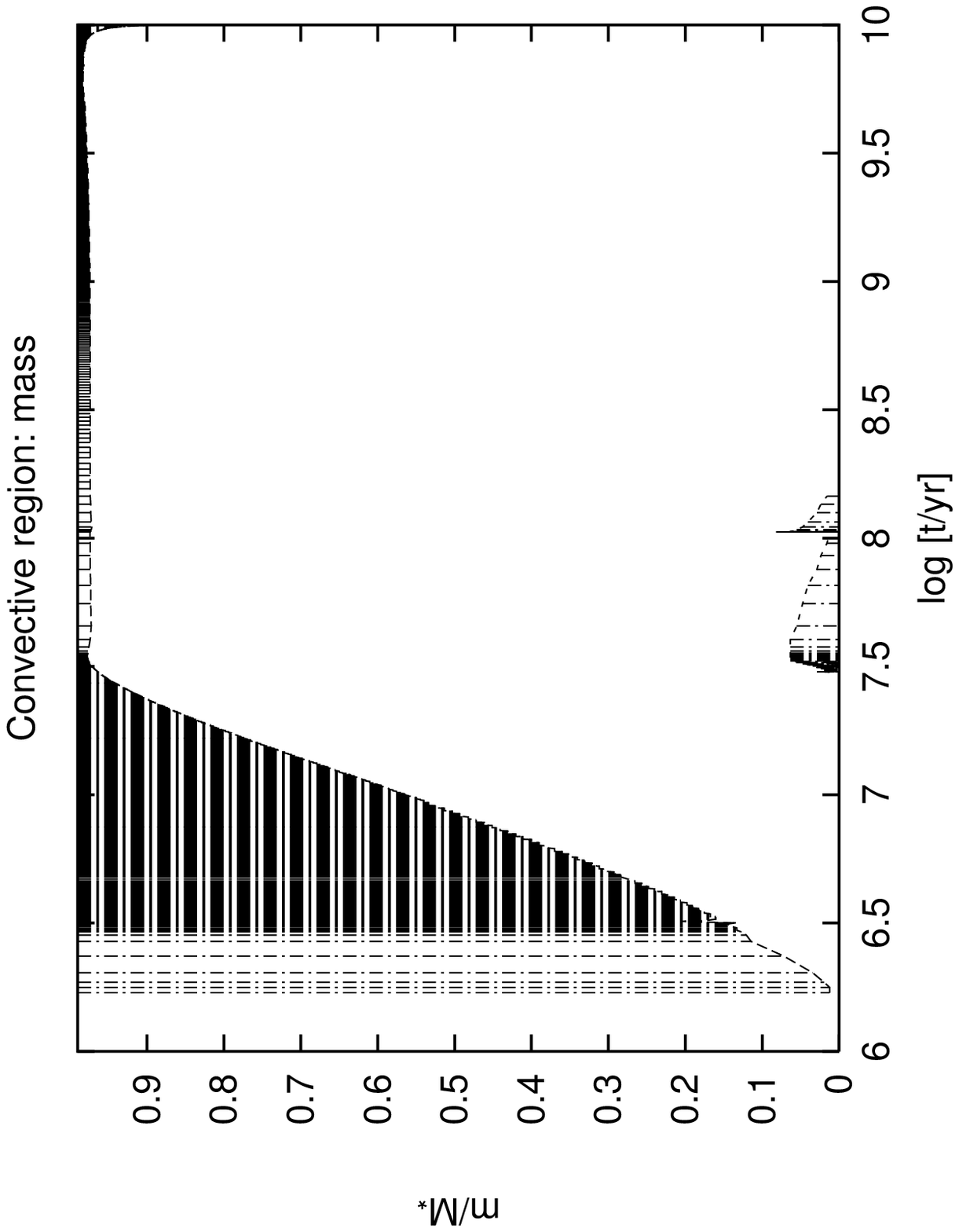}}}
\caption{Extent (in mass units) of the solar convection zone as a function
         of time. Note the transition ($\log t \approx 7.5$) from
         deep-reaching 
         PMS convection to shallow surface convection on the main-sequence. 
                } \label{Fmt}
\end{figure}
\begin{figure}[p]
\centering
\epsfxsize=0.7\textwidth
\mbox{\epsffile{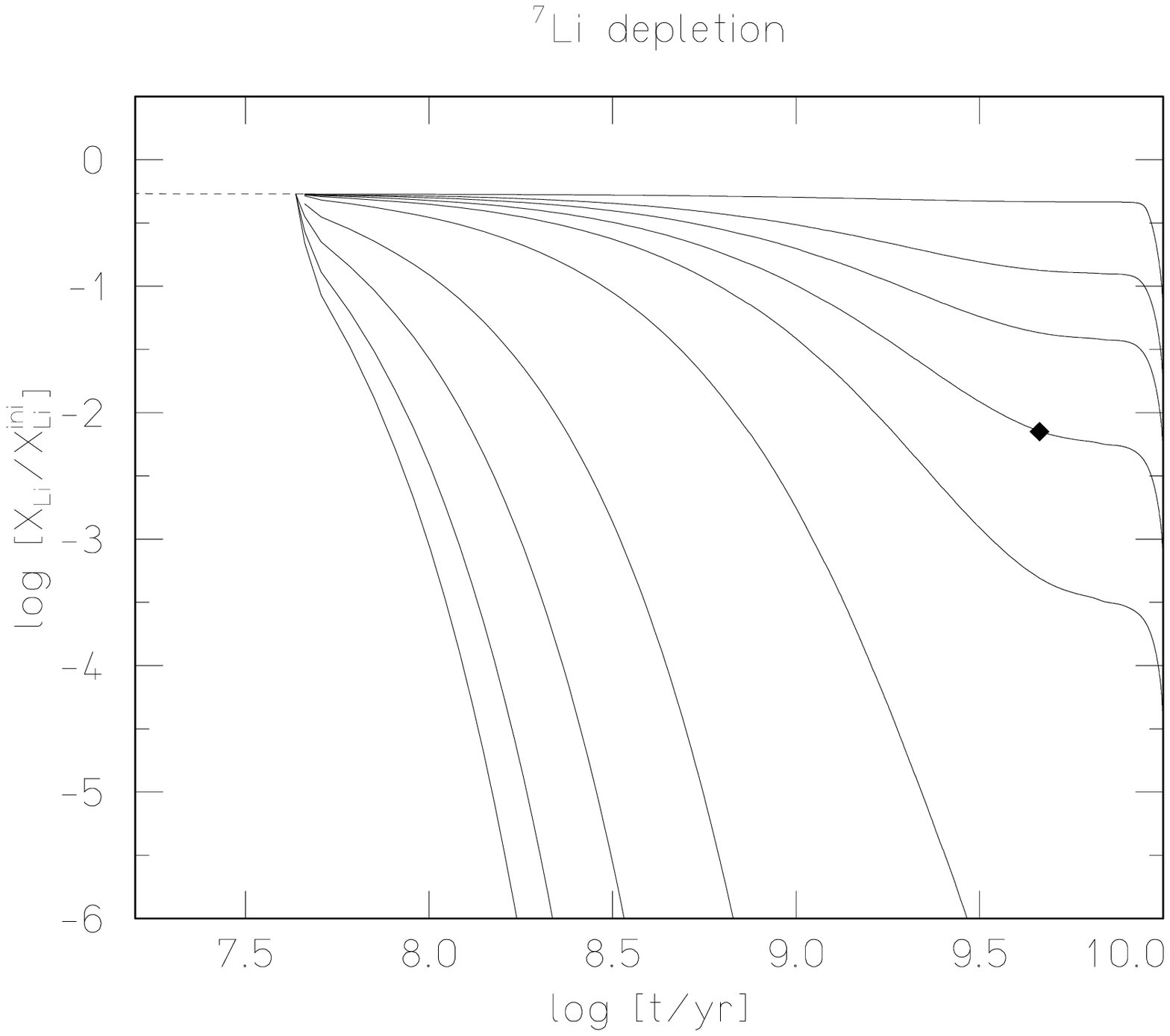}}
\caption{Lithium depletion vs.\ time. The dashed line refers to the
(late) PMS phase, whereas 
the set of solid curves corresponds to different values 
of the mixing efficiency parameter,
$f$ = 0.02, 0.05, 0.06, 0.07, 0.08, 0.10, 0.15, 0.20, 0.26 and 0.31
(from top to bottom), applied during the main-sequence evolution. The diamond
refers to the solar depletion of -2.15 dex.} \label{Fli7depl}
\end{figure}
\begin{figure}[p]
\centering
\epsfxsize=0.7\textwidth
\mbox{\epsffile{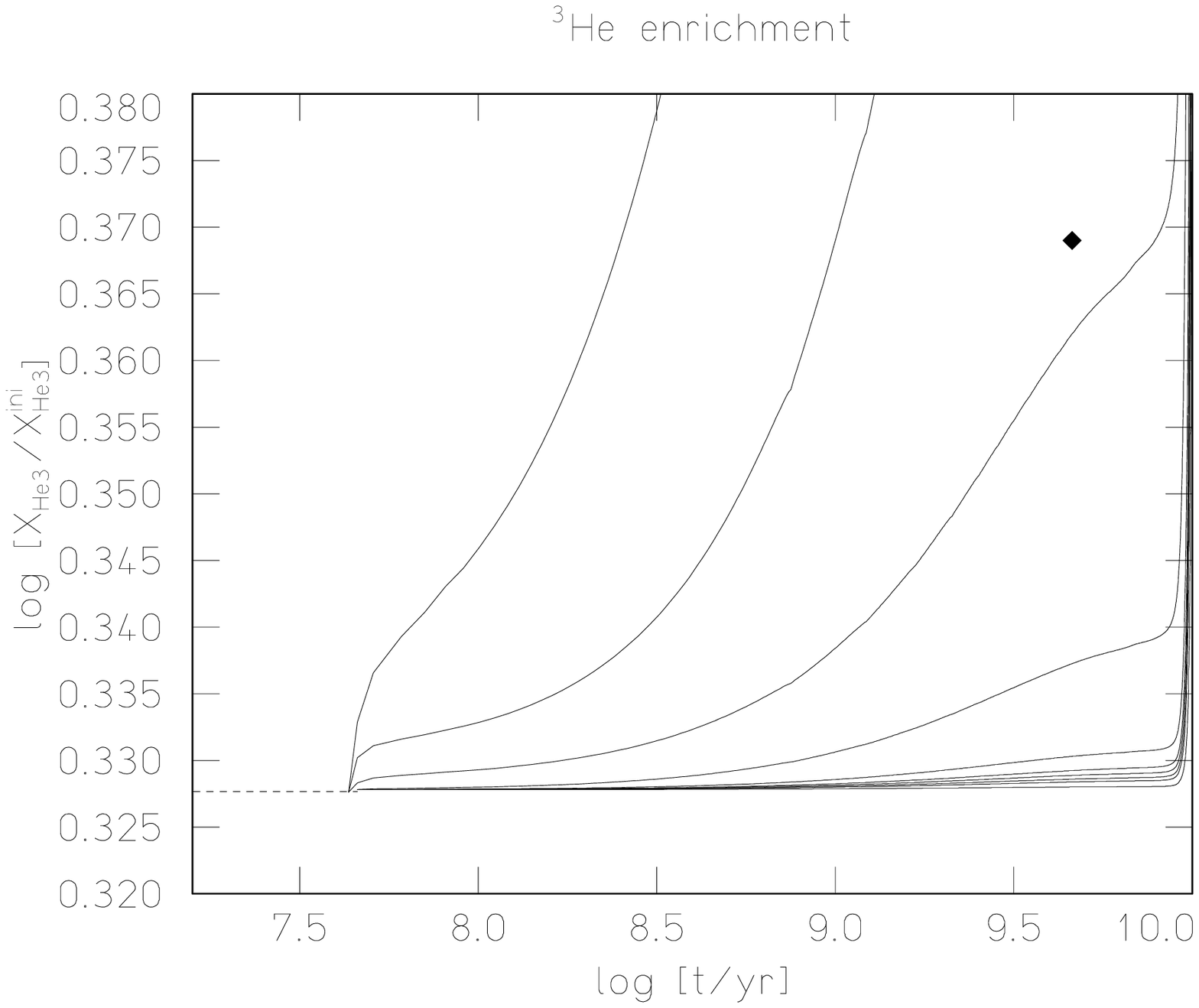}}
\caption{$^{3}$He enrichment vs.\ time. The dashed line refers to the
(late) PMS phase, whereas 
the set of solid curves corresponds to different values 
of the mixing efficiency parameter,
$f$ = 0.02, 0.05, 0.06, 0.07, 0.08, 0.10, 0.15, 0.20, 0.26 and 0.31
(from bottom to top), applied during the main-sequence evolution.
The diamond indicates a 10\% increase during the solar main-sequence
evolution (see text).
} \label{Fhe3enrich}
\end{figure}
Fig.~\ref{Fmt} shows the time evolution of the mass of the convection zone in 
a $1 M_\odot$ model during the pre-main-sequence (PMS) and main-sequence
evolution. Each vertical dash-dotted line corresponds to one stellar model
(for most of the time the models are so closely spaced that a continuous band
appears). 
The deep-reaching convection of the PMS evolution corresponds to a track in
the HRD  starting at the Hayashi limit. There is a well-defined transition to
the shallow, low-mass surface convection zone of the main-sequence star
($\log t \approx 7.5$).

Fig.~\ref{Fli7depl} shows the predicted depletion of lithium during the
main-sequence life 
of a $1 M_\odot$ star ({\em solid lines}).
The set of solid curves corresponds to different values 
of the mixing efficiency parameter,
$f$ = 0.02, 0.05, 0.06, 0.07, 0.08, 0.10, 0.15, 0.20, 0.26 and 0.31.
(from top to bottom).
The present solar value, -2.15, is represented by a diamond, and
is close to the $f$ = 0.07 curve.
The PMS value has been adjusted to -0.3 dex according to observations
of young open clusters (Sect.\,\ref{modcalc}).

Fig.~\ref{Fhe3enrich} illustrates the steady dredge-up of $^3$He,
an intermediate product of 
the p-p chain that has accumulated around $m$\,$\approx$\,$0.6 M_\odot$
during the
first few Gyr of main-sequence evolution. The different curves have the
same meaning as in Fig.~\ref{Fli7depl}, 
but in this case $f$ increases upwards.\,The ZAMS\,value of $^3$He is the sum
of primordial  $^3$He and D, the latter being converted into $^3$He during  
D\,burning in the PMS phase.

\citeauthor{BGM90} \shortcite{BGM90} have used
solar-wind data to constrain the enrichment of photospheric $^3$He 
from the ZAMS to the present. They derive an upper limit of 
10 to 20\,\% and emphasize that this isotope is a sensitive tracer for 
mixing processes. The amount of mixing predicted by our $f = 0.07$\,model 
(which reproduces the observed Li depletion) is less than 1\,\%, 
compatible with their upper limit. We notice that a 10\,\% increase 
of $^3$He
would imply a mixing efficiency of $f \approx 0.2$ (see  Fig.\,4)
which, according to our model, leads to total destruction of Li.
\section{Conclusions}
We have shown that {\em slow mixing}, a diffusion process related to extended
convective overshoot that 
operates in the almost radiative layers underneath a surface convection zone, 
may lead to substantial depletion of lithium during the main-sequence
evolution of a $1 M_\odot$ star. 
The mixing efficiency parameter required to reproduce the observed
{\em depletion of lithium in the Sun}, $f \approx 0.07$, is intermediate
between the parameter range of $0.25 \pm 0.05$ inferred by
\citeauthor{FLS96} \shortcite{FLS96} from
hydrodynamical models of the shallow surface convection of
main-sequence A stars and the value of $f=0.02$ derived empirically by
\citeauthor{HBSE97} \shortcite{HBSE97} from stellar evolution calculations
for core and deep envelope convection.

We believe that the existence of an exponentially decaying velocity field 
below a surface convection zone is a general feature of overshoot.
Although the results of the simulations for A-type stars can certainly not 
be readily applied to the solar case, the basic situation seems not entirely
different: as in A-type stars, a substantial fraction of the total flux is 
in fact carried by radiation in the lower part of the solar convection zone.
In this study we have attributed the depletion
of Li exclusively to extended overshoot. If other mixing processes
should prove to contribute as well, the efficiency of overshoot
will be smaller than derived here, i.e. $f < 0.07$.
However, in view 
of the success of the simple mixing model presented here we feel 
that the potential importance of slow mixing in the context of stellar 
convection deserves further study.

\end{document}